\documentclass[twocolumn,trackchanges,dvipsnames]{aastex6}
\RequirePackage{lineno}
\usepackage{amsmath}
\usepackage{amssymb}
\usepackage{amsthm}
\usepackage{natbib,aasdefs,url,bm}

\usepackage[colorlinks=true]{hyperref}
\hypersetup{
    colorlinks=true,
    linkcolor=RoyalBlue,
    filecolor=magenta,      
    urlcolor=orange,
    citecolor=BrickRed
}

\newcounter{ichi}
\newcounter{ni}
\newcounter{san}
\newcounter{yon}
\def\be{\begin{equation}}
\def\ee{\end{equation}}
\def\ba{\begin{eqnarray}}
\def\ea{\end{eqnarray}}

\def\aap{\ {A\&A}\ }

\def\apjl{\ {ApJL}\ }
\def\apjs{\ {ApJS}\ }

\shorttitle{}
\shortauthors{Kheirandish and Murase}

\begin{document}

\title{Detecting High-Energy Neutrino Minibursts from Local Supernovae\\ with Multiple Neutrino Observatories}

\author{Ali Kheirandish\altaffilmark{1}}
\altaffiltext{1}{Department of Physics \& Astronomy; Nevada Center for Astrophysics, University of Nevada, Las Vegas, NV 89154, USA}
\author{Kohta Murase\altaffilmark{2,3,4}}
\altaffiltext{2}{Department of Physics; Department of Astronomy \& Astrophysics; Center for multi-messenger Astrophysics, Institute for Gravitation and the Cosmos, The Pennsylvania State University, University Park, PA 16802, USA}
\altaffiltext{3}{School of Natural Sciences, Institute for Advanced Study, Princeton, NJ 08540, USA}
\altaffiltext{4}{Center for Gravitational Physics and Quantum Information, Yukawa Institute for Theoretical Physics, Kyoto University, Kyoto, Kyoto 606-8502, Japan}
\altaffiltext{5}{School of Natural Sciences, Institute for Advanced Study, Princeton, New Jersey
08540, USA}
\begin{abstract}
Growing evidence from multi-wavelength observations of extragalactic supernovae (SNe) has established the presence of dense circumstellar material in Type II SNe. Interaction between the SN ejecta and the circumstellar material should lead to diffusive shock acceleration of cosmic rays and associated high-energy emission. Observation of high-energy neutrinos along with the MeV neutrinos from SNe will provide unprecedented opportunities to understand unanswered questions in cosmic-ray and neutrino physics. 
We show that current and future neutrino detectors can identify high-energy neutrinos from an extragalactic SN in the neighborhood of the Milky Way. We present the prospects for detecting high-energy neutrino minibursts from SNe in known local galaxies, and demonstrate how the network of multiple high-energy neutrino detectors will extend the horizon for the identification of high-energy SN neutrinos. We also discuss high-energy neutrino emission from SN 2023ixf. 
\end{abstract}


\maketitle

\section{Introduction}
Neutrinos from supernova (SN) 1987A were the first observed astrophysical neutrino signal beyond the solar system~\citep{Kamiokande-II:1987idp,Bionta:1987qt}. 
That observation tremendously revolutionized our understanding of the collapse of massive stars and highlighted the fundamental role of neutrinos~\citep{Raffelt:1990yu,Kotake:2005zn,Janka:2016fox}. 
More than two decades later, observation of high-energy cosmic neutrinos by the IceCube Neutrino Observatory~\citep{Aartsen:2013bka, Aartsen:2013jdh,Aartsen:2014gkd,Aartsen:2020aqd} demonstrated the abundant energy carried by neutrinos in the high-energy universe and revealed the greater than the expected role of hadronic interactions in nonthermal processes.

Neutrinos from core-collapse SNe (CCSNe) are crucial probes of explosive phenomena at the deaths of massive stars and neutrino physics. High-energy neutrinos are produced through hadronic processes of cosmic rays (CRs), facilitated by the interaction between the SN ejecta and dense circumstellar material (CSM). Together with the MeV neutrinos from SNe, the high-energy neutrino emission offers an unprecedented opportunity for studying particle acceleration, as well as fundamental properties of neutrinos and physics beyond the Standard Model in the neutrino sector, and \cite{Murase:2017pfe} proposed Galactic CCSNe as excellent ``multi-energy'' neutrino sources.

Transient astrophysical phenomena present the most promising channel for identifying the origin of high-energy cosmic neutrinos. They benefit from a lower number of background events, therefore, enhancing the likelihood for significant observations via multi-messenger searches. So far, neutrino-triggered follow-ups have presented tantalizing candidates of two transient neutrino sources. The first compelling evidence was suggested by the coincidence of a high-energy neutrino with a flaring blazar~\citep{IceCube:2018cha,IceCube:2018dnn}. 
Subsequently, two high-energy neutrinos were found coincident with tidal disruption events~\citep{Stein:2020xhk,Reusch:2021ztx}. 
These developments have demonstrated the power of time-domain neutrino astronomy. See \citet{Murase:2019tjj} for a review about transient sources of high-energy neutrinos.

The promise of high-energy neutrino detection from SNe was realized after growing observations of pre-explosion mass losses or envelope inflation before the explosion in extragalactic SNe~\citep{Smith:2014txa,Smith:2007cb,Immler:2007mk,Miller:2008jy,Ofek:2013mea,Margutti:2013pfa,Margutti:2016wyh,Yaron:2017umb,Morozova:2017hbk}. 
\citet{Murase:2017pfe} first pointed out that with such ``confined'' CSM (CCSM) even ordinary Type II (II-P/II-L/IIb) SNe are detectable with high statistics by current neutrino detectors. This scenario has been strongly motivated by recent observations with flash spectroscopy and modeling of early light curves~\citep[e.g.,][]{2020MNRAS.496.1325B}, and the model parameters can be determined by multi-wavelength observations~\citep{Murase:2018okz}. The SN shock eventually breaks out from the CSM or envelope, and shallow density profiles of the CSM inevitably lead to the formation of collisionless shocks and the onset of diffusive shock acceleration (DSA) of ions~\citep{Murase:2010cu,Katz:2011zx,Murase:2013kda}. 
This scenario is different from models assuming neutrino production in the choked jets~\citep[e.g.,][]{Murase:2013ffa, Senno:2015tsn, He:2018lwb, Fasano:2021bwq, Guetta:2019wpb}. Our scenario does not involve jets whose existence is not guaranteed especially for Type II SNe, and particle acceleration in powerful choked jets is limited when the shock is radiation mediated \citep{Murase:2013ffa}. 
Depending on SN types, within $\sim1-100$~days after the MeV thermal burst of neutrinos, high-fluence neutrino emission from CCSNe is anticipated. Such observation would provide the first multi-energy neutrino source~\citep{Tamborra:2018upn}.

\begin{figure}[t!]
    \centering
    \includegraphics[width=\columnwidth]{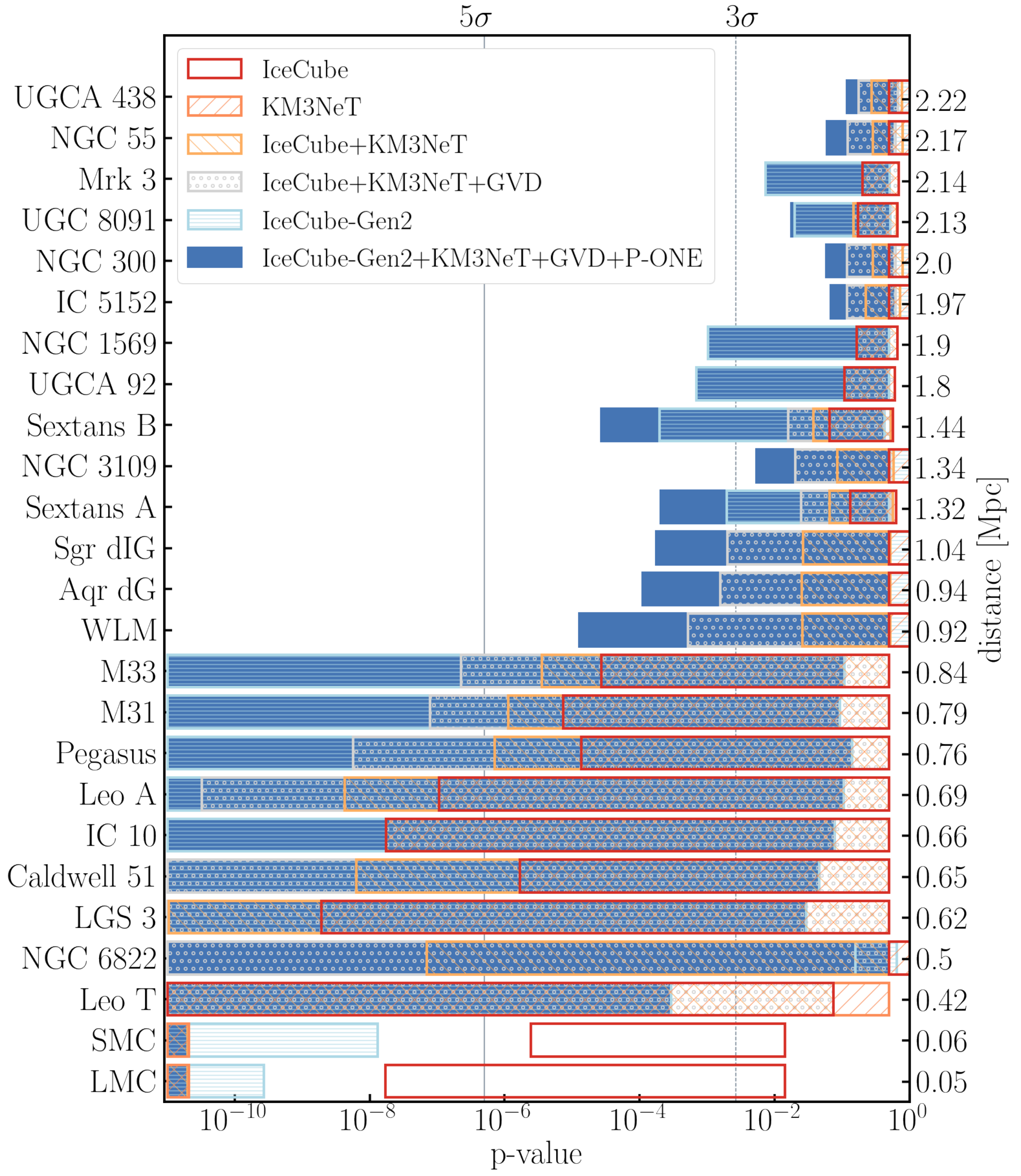}
    \caption{Prospects for identifying neutrino emission from SNe II with CCSM in local galaxies within a distance of 2~Mpc. Here, we show the expected p-values for $D_*$ ranging from 0.01 (right edge) to 1 (left edge). The list of galaxies is complied from~\citet{Nakamura:2016kkl}. We present the expected p-values for IceCube (red), KM3NeT (orange hatched), joint IceCube and KM3NeT (yellow), combined IceCube, KM3NeT, and Baikal-GVD (light blue), IceCube-Gen2 (hatched blue), and combined analysis of IceCube-Gen2, KM3NeT, GVD, and P-ONE (dark blue).}
    \label{fig:pval_local}
\end{figure}

The rate of CCSNe in the Milky Way is estimated to be $\sim$ 3 per century~\citep{Adams:2013ana}.
Since SN 1987A, no SN has been observed with neutrino detectors. Hence looking for extragalactic SNe may be a faster way to detect neutrinos, and the star-formation rate in local galaxies is higher than in the continuum limit~\citep{Ando:2005ka}. 
In the meantime, optical surveys have so matured and we can monitor the sky with high cadence not to miss local SNe. 
Given these recent developments, it is crucial to evaluate high-energy neutrinos from nearby CCSNe. 
Currently, IceCube is the only cubic kilometer-scale operational detector that is most sensitive to the sources in the Northern Sky, thanks to Earth's shielding the atmospheric muon background. In the near future, the commissioning of KM3NeT in the Mediterranean~\citep{Adrian-Martinez:2016fdl}, and GVD in the lake Baikal~\citep{Avrorin:2019vfc} will provide coverage to the Southern Sky as well increasing the overall coverage. The addition of P-ONE~\citep{Agostini:2020aar} would enhance the overall coverage in the Northern hemisphere. The next generation detector, IceCube-Gen2, will additionally expand the exposure and provide substantial sensitivity to search for high-energy neutrinos~\citep{Aartsen:2014njl}. 
Here, we present an update for the detectability of high-energy neutrinos from extragalactic CCSNe. 
The modeling involves the enhancement of the CCSN rate in local galaxies, improvement of $pp$ interactions at lower energies, and more systematic studies on CSM densities. Large CSM parameters, which are inferred by recent observations~\citep[e.g.,][]{2020MNRAS.496.1325B}, can enhance the high-energy neutrino flux by many orders of magnitude, which makes the detection of extragalactic CCSNe feasible. Consequently, we show that the horizon for observation of high-energy neutrinos from CCSNe will be extended beyond the Milky Way with the network of current and future neutrino telescopes. Figure~\ref{fig:pval_local} shows the expected p-value for identifying high-energy neutrinos from interacting SNe occurring in local galaxies, in the range of 2 Mpc, with IceCube and a network of future neutrino telescope. While a SN in nearby galaxies such as LMC and SMC could be identified with a single detector, the significance of the identification of high-energy neutrinos from SNe will be highly enhanced with joint analysis and observation of multiple detectors.
For instance, joint observation of the network of current and upcoming neutrino telescopes would result in reaching $5\sigma$ significance for SNe that occur in galaxies such as Calwell 51, Pegasus, and Andromeda (M31).

\section{High-Energy Neutrinos from CCSNe}
Recent observations have indicated that CCSNe commonly have CCSM with a size of $R_w\sim{\rm a~few}\times10^{14}-10^{15}$~cm~\citep{2020MNRAS.496.1325B,Terreran:2021hfc,Chugai:2022fcz}. DSA should operate as in SN remnants, which predicts CR energy fractions of $\sim10-20$\% and spectral indices of $s_{\rm cr}\sim2$, being supported by numerical simulations~\citep{1977ICRC...11..132A,1977DoSSR.234.1306K,1978MNRAS.182..147B,1978ApJ...221L..29B,Schure:2012du,Caprioli:2013dca,Caprioli:2019xhb}. 
This situation is different from high-energy neutrino emission from jets, in which CR acceleration is highly uncertain~\citep{Meszaros:2001ms,Ando:2005xi,Murase:2006mm,Murase:2013ffa}. Neutrino detection is crucial, because neutrinos can reveal the onset of DSA through the transition from collisional/radiation mediated to collisionless shocks~\citep{Murase:2010cu,Katz:2011zx}. 

Following \citet{Murase:2017pfe}, we consider a wind density profile of $\varrho_{\rm cs} = D r^{-2}$ up to $R_w$, where the coefficient $D$ is related to mass loss rate ($\dot{M}$) and wind velocity ($V_w$) as
\begin{equation}
    D = \frac{\dot{M}}{4\pi V_w},
\end{equation}
where we define $D_* = D/(5\times 10^{16} \rm \, g \, cm^{-1})$. For ordinary SNe II, we incorporate large values of $D_*$ as indicated by the modeling for the mass and extent of the dense CSM in SNe II~\citep{Yaron:2017umb,Morozova:2017hbk,Forster:2018mib,Bruch:2020jcr,Bruch:2022aqd,2020MNRAS.496.1325B,Terreran:2021hfc,Chugai:2022fcz}.
As in \citet{Murase:2017pfe} , we take into account the details of shock evolution following the self-similar solution \citep{1982ApJ...258..790C,1985Ap&SS.112..225N,Moriya:2013hka}, which is consistent with results of the numerical simulations reported by~\citet{Tsuna:2023zwm,Tsuna:2019srj}.
Throughout this work, we consider the allowed range from $D_*=0.01$ (as indicated from SN 2013fs) to $D_*=1$~\citep[as found in, e.g.,][]{2020MNRAS.496.1325B,Chugai:2022fcz}. For more details on the updated neutrino fluence for ordinary SNe II, see Sec.~\ref{flux} in Appendix.
For SNe IIn, we adapt $D_*=1$ as was assumed in the previous work~\citep{Murase:2017pfe}.

If a long-awaited Galactic SN happens, CR acceleration and associated high-energy neutrino emission occurs within $\sim10$ days after the MeV neutrino burst~\citep{Murase:2017pfe}. For $D_*=1$, a high statistics signal is expected for both SNe II and IIn in IceCube with more than $10^5$ and $10^7$ events, respectively. Although the level of the expected number of neutrino events rapidly drops for SNe outside the Milky Way, the predicted horizon for observation of an interacting SN II, is $\sim 1$~Mpc for IceCube, which will be extended to more than 5~Mpc in IceCube-Gen2. 
For SNe IIn, the horizon extends to $\sim7$~Mpc in IceCube and $\sim15$~Mpc for IceCube-Gen2~\citep{Murase:2010cu, Petropoulou:2017ymv}.

\section{Prospects for Observations with Multiple Neutrino Detectors}
We explore the prospects for the identification of high-energy neutrinos from ordinary SNe II and IIn in current and future neutrino telescopes. 
We first focus on IceCube, as the current operating km$^3$ scale neutrino telescope, and then will discuss the prospects for the neutrino telescope being deployed in the Mediterranean sea, KM3NeT, as well as the one in the lake Baikal, Baikal-GVD. 
Finally, we present the estimates for the next generation neutrino detector, IceCube-Gen2, at the South pole. Here, we calculate the number of neutrino events with energies greater than 100~GeV and follow the prescription described in~\citet{CMS-NOTE-2011-005} to calculate the p-value for the identification of events in each detector. For more details, see Appendix. 

\begin{figure}[t]
    \centering
    \includegraphics[width=\columnwidth]{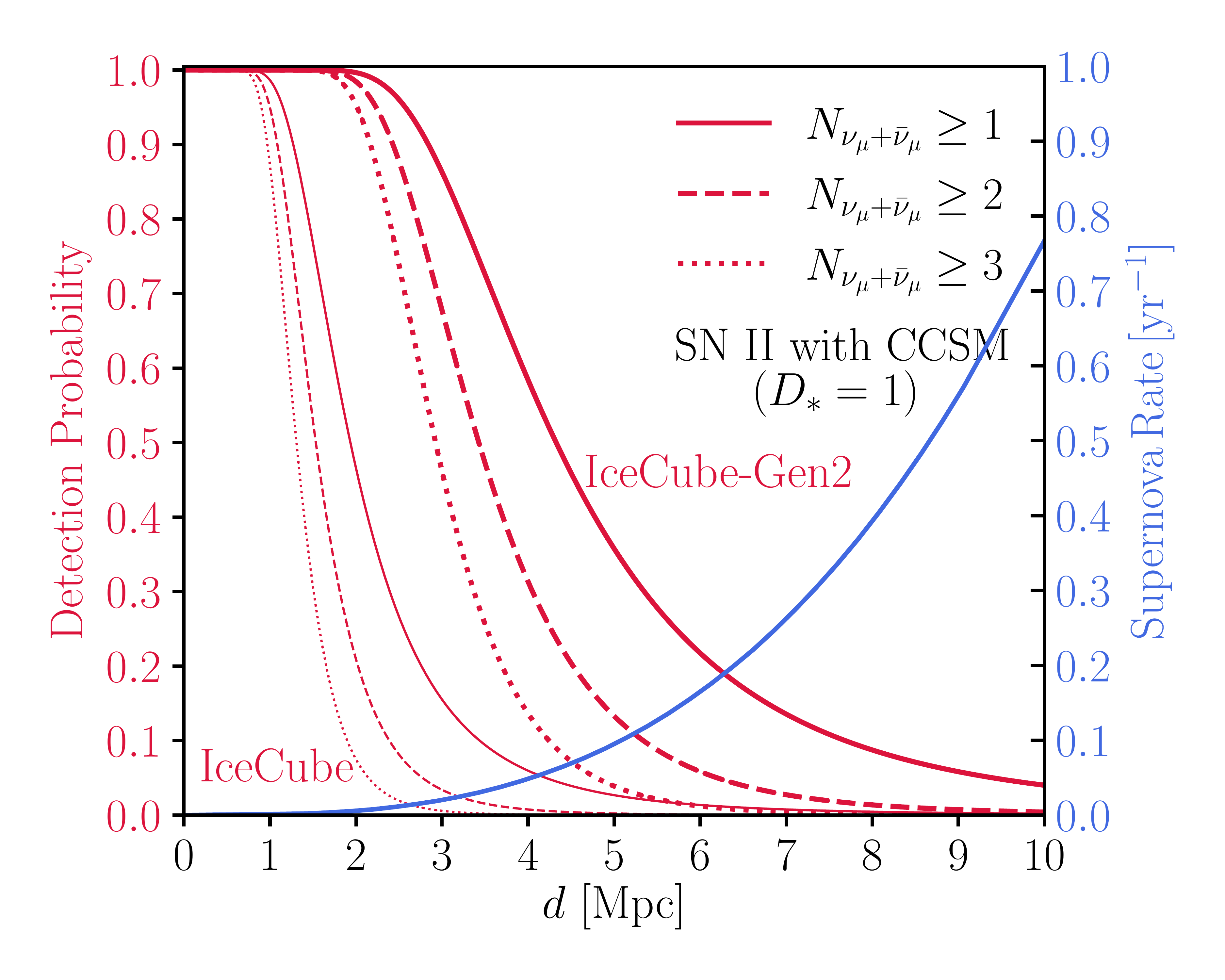}
    \caption{Detection probability for high-energy neutrinos from SN IIe with CCSM in IceCube and IceCube-Gen2, assuming events in the Northern sky. We show the probability of observation of at least 1 (solid), doublet (dotted), and triplet (dashed) events for IceCube (thin) and IceCube-Gen2 (thick). Here, we consider a CR spectral index of $s_{\rm cr}=2.0$ and the optimistic scenario ($D_*=1$). The blue line shows the expected CCSN rate based on~\citet{Nakamura:2016kkl} with the relative fraction of SNe II with CCSM (60\% of CCSNe) taken into account.
    See text for details.}
    \label{fig:sn_iip_det_prob}
\end{figure}

{\em -- Ordinary SNe II:} Type II SNe are the dominant component of the CCSNe~\citep{Graur:2016lca}. The frequency of SNe II-P is reported to constitute from $\sim$ 50\%~\citep{Smith:2010vz,2011MNRAS.412.1441L} of the CCSNe.  Here, given that even SNe II-L/IIb may have CCSM, we choose a moderate value and set the frequency of SNe II with CCSM to 60\% of CCSNe.
The more frequent occurrence of interacting SNe II increases the likelihood of its identification. 
Figure~\ref{fig:events-sn-ii} illustrates the expected number of events from SNe II with CCSM in IceCube and IceCube-Gen2. As mentioned earlier, in light of the recent studies suggesting larger $\dot{M}$ for SNe II~\citep{2020MNRAS.496.1325B,Terreran:2021hfc,Chugai:2022fcz,WH15}, we consider $D_* =0.1$ and $D_* =1.0$~\citep[see Table~2 of][]{2020MNRAS.496.1325B}, in which for the first time we show that the likelihood for observation of ordinary SNe in neutrino telescopes is enhanced more than an order of magnitude.
Incorporating larger values of $D_*$, we find that the time-integrated neutrino flux from a Galactic SN (10~kpc) exceeds $10^4 \, \rm GeV \, cm^{-2}$ in the range of 100 GeV to 1 PeV, which is the effective range of energies current generation of neutrino telescopes operate. In examining the prospects for the identification of this type, therefore, we provide the range of p-values according to different values of $D_*$. The optimal time for SNe II-P, where the signal-to-background is maximal, depends on the assumed value of $D_*$~\citep{Murase:2017pfe}, that is $t_{\rm max} = 10^{5.8}$ s for $D_*=0.01$, $10^6$ s for $D_*=0.1$, and finally $10^{6.6}$ s for $D_*=1$. 

It is important to consider the fact that the Milky Way resides in the structured region of the universe, and the CCSN rate in local galaxies is larger~\citep{Nakamura:2016kkl}. Figure~\ref{fig:pval_local} shows the prospects for the identification of interacting SNe II in local galaxies for $D_*$ ranging from 0.01 to 1. We begin with those in IceCube and follow with adding upcoming telescopes. Because of the statistical limitations, we assume a floor of $10^{-11}$ for the p-value as the generation of larger event samples was computationally expensive and unnecessary. The closest local galaxies are Large and Small Magellanic Clouds (LMC and SMC). Despite being located in the Southern sky, the neutrino fluence for a SN II with CCSM at their distance still can provide a substantial number of muon neutrinos in IceCube, which can make the identification feasible with a significance better than $3\sigma$. Within 2 Mpc, a good fraction of local galaxies reside in the Northern sky. Therefore, providing IceCube with a good likelihood for identification with even moderate values of $D_*$. As the sources become more distant, the significance for their identification would drop below $3\sigma$. The addition of KM3NeT in the Northern hemisphere will enhance the sensitivity for the sources in the Southern sky. For LMC and SMC, KM3NeT alone can surpass the significance achievable in IceCube. Identification at the discovery level is even possible for Leo T. Joint observations of IceCube and KM3NeT will enhance the sensitivity and can bestow $5\sigma$ observation for an additional galaxy, Caldwell 51, while enhancing the significance for identification for dozen more sources better than $3 \sigma$. The addition of Baikal-GVD will boost the sensitivity for identification of SNe in the Southern sky and will further provide better significance. In particular, the joint observation of IceCube, KM3NeT, and Baikal-GVD enables detection at $5\sigma$ for a SN II in Pegasus and M31.

Another major improvement in the prospects for identification will be provided by the commissioning of IceCube-Gen2 that would realize nearly ten times the volume of IceCube. Identification with better than $5\sigma$ is feasible for the majority of sources near 1 Mpc, and further than that, the addition of IceCube-Gen2 and combined analysis of the present detectors can provide evidences for neutrino emission from CCSNe to $\sim2$~Mpc.

The key development in the identification of high-energy neutrinos from SNe relies on either the larger effective volume brought by the next-generation neutrino detectors or the combined search by a network of current cubic kilometer-scale detectors.
Fig.~\ref{fig:gen2-pval} in Appendix shows the estimated p-values for identification of SNe II with CCSM in IceCube-Gen2 up to a distance of 10 Mpc. 

We also estimate the detection probability of multiplets from SNe II with CCSM as a function of distance for the current and next generation of neutrino telescopes. Figure~\ref{fig:sn_iip_det_prob} shows the Poisson probability for identification of at least a singlet, doublet, and triplets. For $D_*=1$, the probability of detecting a doublet is close to 50\% around 4 Mpc. With the growing number of galaxies and a higher rate of SNe II with CCSM, this estimate implies that identification of such interacting SNe will be likely with the operation of future neutrino telescope. For comparison, we also show the expected rate for SNe II with CCSM in Fig.~\ref{fig:sn_iip_det_prob}. This rate is analytically estimated from the star-formation rate. It is worth noting that the observed rate for CCSNe is, in fact, higher than this analytical estimate (see~\citet{Horiuchi:2013bc} for details), which makes over projected prospects relatively conservative. 

\begin{figure}[t!]
    \centering
    \includegraphics[width=\columnwidth]{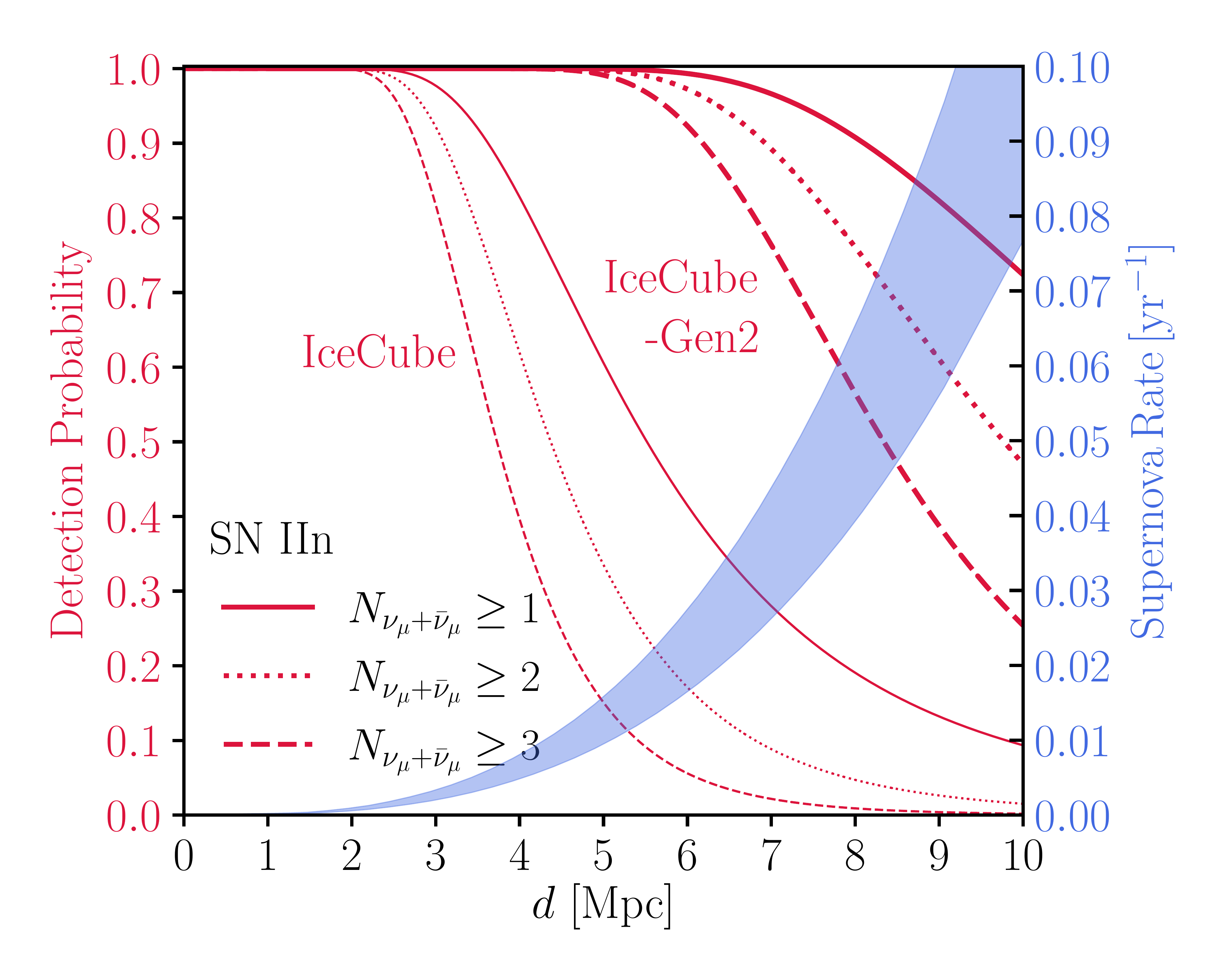}
    \caption{Detection probability for high-energy neutrinos from SN IIn in IceCube and IceCube-Gen2, assuming events in the Northern sky. We show the probability of observation of at least 1 (solid), doublet (dotted), and triplet (dashed) events for IceCube (thin) and IceCube-Gen2 (thick). Here, we consider a CR spectral index of $s_{\rm cr}=2.0$. The expected rate of SNe IIn is also shown with the blue band. The width of the band shows the uncertainty associated to the measured rate.}
    \label{fig:sn-iin-prospects}
\end{figure}  
{\em -- SNe IIn:} For Type IIn SNe, we consider the optimal time ($t_{\rm max}$) of $10^{7.5}$ s. Thanks to the high level of the neutrino fluence, the horizon for expecting one event from a SN IIn in IceCube extends to $\sim$ 7~Mpc, provided that the source is in the Northern hemisphere. 
Similarly, KM3NeT and Baikal GVD can detect at least a muon neutrino from a SN in the Southern sky, though the number might slightly vary depending on the visibility of the source. 
Figure~\ref{fig:events-sn-ii} in Appendix shows the expected number of neutrino events from SN IIn as a function of distance. Observation with enough statistical significance (i.e., 3$\sigma$) for a single source is possible for IceCube until $\sim$ 2 Mpc. IceCube-Gen2 will nearly double this horizon to more than 4 Mpc.
See Fig.~\ref{fig:sn-iin-prospects} in Appendix for more details. This significance horizon is solely defined for distinguishing signal events from the background, regardless of multi-messenger information. When cross-correlating neutrinos with optical and/or gamma-ray signals, it is beneficial to estimate the likelihood for identification of multiplet of events. As such, we estimate the Poisson probability of having at least a singlet, doublet, or triplet events from SNe IIn in IceCube and IceCube-Gen2. 
Fig.~\ref{fig:sn-iin-prospects} shows the detection probability for each multiplet expectations. Thanks to the high level of fluence for SN IIn, the probability of having more than two events is $\sim$ 50\% for a SN at 10 Mpc. 
We also show the projected rate for SNe IIn. The caveat here, however, is that the Type IIn SNe are subdominant components of the CCSNe. Thus, despite the high level of emission, their rare occurrence does not yield high feasibility for their observation. 

\begin{figure*}[t]
    \centering
    \includegraphics[width=\columnwidth]{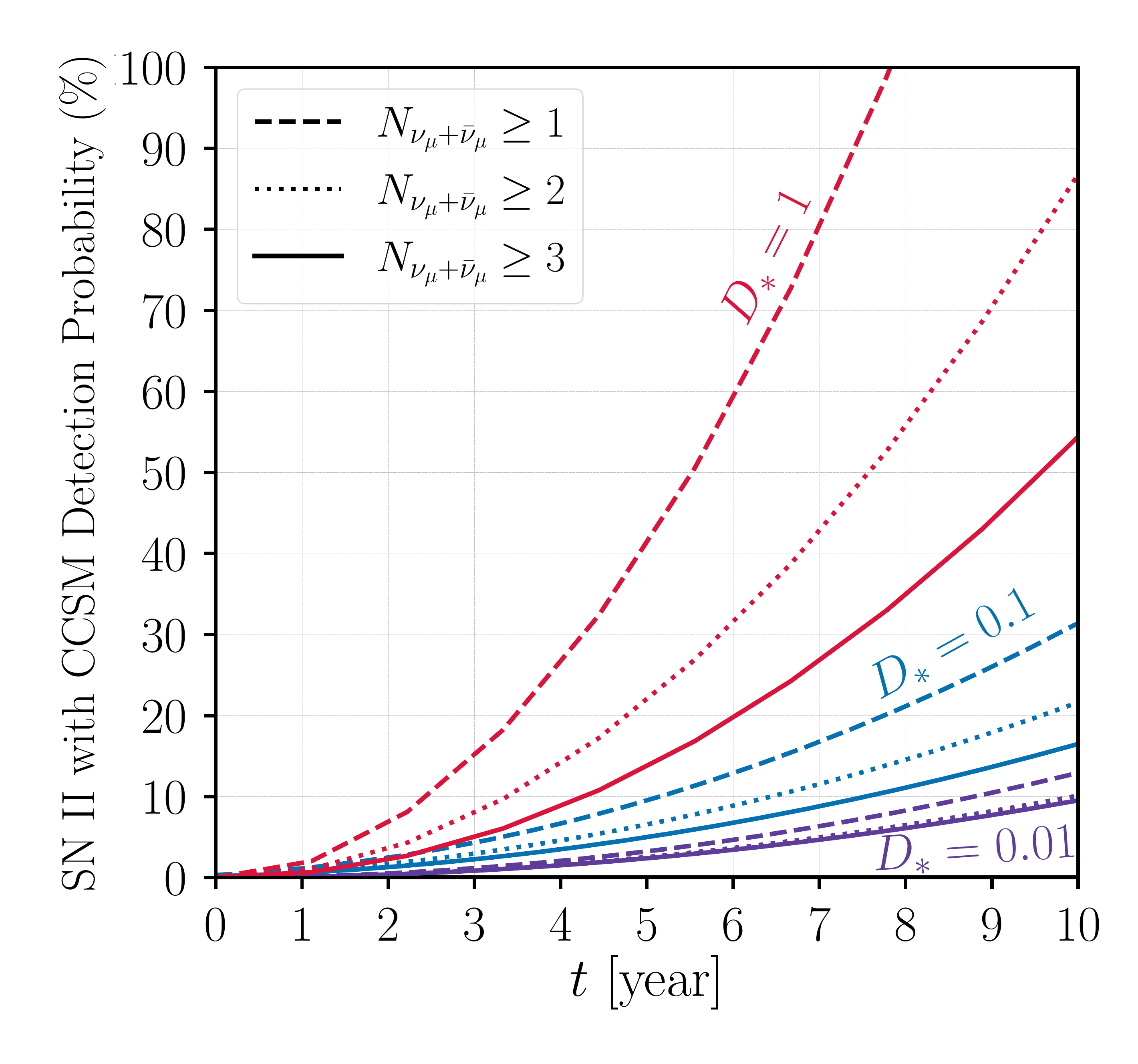}
    \includegraphics[width=\columnwidth]{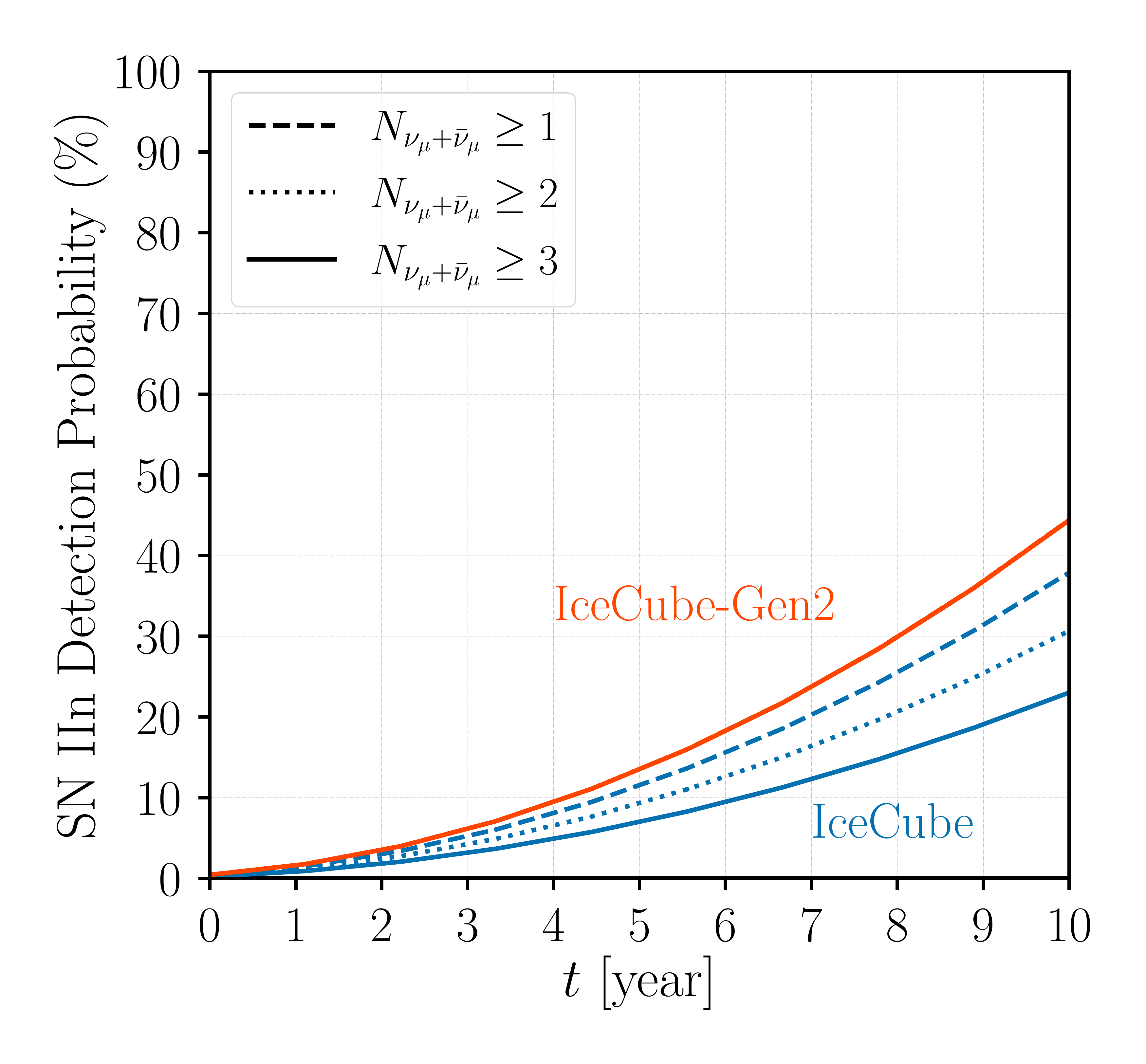}
    \caption{Probabilities of detecting singlet (solid), doublet (dotted), and triplet (dashed) neutrino events from SNe II with CCSM and SNe IIn up to a distance of 5 Mpc in the next generation of neutrino telescopes. Here, we consider the list of galaxies provided by \cite{Nakamura:2016kkl}.}
    \label{fig:gen2-dtection-prob}
\end{figure*}

We should note that here we are considering a varying velocity for the ejecta, in contrast to other estimates of the neutrino emission from SN IIn that assume constant velocity~\citep{Petropoulou:2016zar,Zirakashvili:2015mua}. The variable velocity, in fact, enhances the likelihood of observation. 

The projected p-values for the identification of high-energy neutrinos from CCSNe in local galaxies reflect the feasibility of distinguishing signal and background if a SN goes off in each of these galaxies. The probability for observation of high-energy neutrinos from CCSNe also is tied to the rate of CCSN in each of these galaxies. Therefore, we calculate the probability for observation of neutrino multiplets from CCSNe in local galaxies according to their reported rate. For this purpose, we estimate the detection probability of at least $N$ neutrinos from a SN at a galaxy of distance $D$, and then, using the CCSN rate in that galaxy as reported by~~\citet{Nakamura:2016kkl}, we calculate the occurrence probability as $P(r) = 1-\exp(r t)$, where $r$ is the local CCSN rate, and $t$ is the observation period. The probability for detection of high-energy neutrinos from CCSNe, therefore, will be the integral of the product of the two probabilities.

Figure~\ref{fig:gen2-dtection-prob} shows the probability for detection of SNe IIn in IceCube and IceCube-Gen2, and SNe II with CCSM in IceCube-Gen2, assuming 10 years of operation. Here, we consider local galaxies up to 5 Mpc. For SNe II with CCSM, the probability ranges from $\sim$ 10\% for conservative values of $D_*$ and reaches unity for optimistic scenarios. Benefiting from a higher rate, even moderate scenarios for interacting SNe project a higher probability for observation with the next generation of neutrino telescopes.

\section{Summary and Outlook}
We presented an updated estimate for high-energy neutrino emission from nearby CCSNe, and demonstrated for the first time that the observation of minibursts of high-energy neutrinos from CCSNe in local galaxies is likely with the commissioning of multiple neutrino telescopes in the near future. The operation of the high-energy neutrino detector network will present a unique and unprecedented opportunity to establish robust probes of confined CSM, shock physics, DSA mechanisms for ions, and neutrino properties. 

Neutrino emission from SNe has been targeted by the transients searches conducted by the IceCube Collaboration~\citep{IceCube:2018omy}. So far, no significant association of high-energy neutrinos with SNe has been reported. The most recent stacking analysis by the IceCube Collaboration~\citep{IceCube:2021oiv} finds limits compatible with the theoretical expectations. The high level of the neutrino fluence at sub-TeV energies could also be targeted with the specialized searches in IceCube and future neutrino telescopes~\citep{IceCube:2020qls}. 
In addition, a recent identification of a CCSN in M101, enhanced motivations for identifying high-energy neutrino emission from core collapse phenomena. Within  the model presented in this study, due to the distance of SN 2023ixf and its II-P nature, less than 0.02 muon neutrinos are expected in IceCube. 
This is in agreement with the absence of signal in the IceCube follow up of the event \citep{IceCubeSN2023}. For more details on the predicted flux and event expectation, see Sec.~\ref{ixf} in Appendix. We note that \citet{Guetta:2023mls} has investigated both low- and high-energy neutrino emission from SN 2023ixf.

In the near future, Hyper-Kamiokande (HK) will enhance the sensitivity to neutrino emission from CCSNe in the sub-TeV range as well as the MeV range. While the the detectability of minibursts from CCSNe with network of high-energy neutrino telescopes is comparable to that of HK for MeV neutrinos  \citep{Nakamura:2016kkl}, in the event of nearby SNe, HK will bring valuable information on the low-energy part of the spectrum that is hard to study with neutrino telescopes in the ice or the sea and will enable us to study CCSNe as multi-energy neutrino neutrino sources as in Galactic SNe \citep{Murase:2017pfe}. Moreover, future searches for high-energy neutrino emission from CCSNe will largely benefit from the advancements in optical surveys brought by the operation of the Rubin Observatory, addition of wide-field time-domain space telescopes such as ULTRASAT \cite{Shvartzvald:2023ofi} along with the current facilities like Zwicky Transient Facility, ASAS-SN, Tomoe-Gozen, and Subaru-HSC. 

\textbf{\textit{Acknowledgements}}
We would like to thank Segev Benzvi and Francis Halzen for their useful comments and suggestions. The authors acknowledge the support from Kavli Institute for Theoretical Physics. This research was supported in part by the National Science Foundation under Grant No.~NSF PHY-1748958. A.K. acknowledges the support from IGC Postdoctoral Award. The work of K.M. is supported by NSF Grant No.~AST-1908689, and KAKENHI No.~20H01901 and No.~20H05852. 

{\em Note added.—} While we were finalizing this project, we became aware of arXiv:2204.03663~\citep{Sarmah:2022vra}. The focus is different and we rely on the dedicated time-dependent kinetic code developed in \citet{Murase:2017pfe}, and there are notable differences regarding the parameters employed and the projected fluxes for both types of SNe considered in this work. In addition, we note that this work was earlier presented at TAUP 2021~\citep{TAUP2022} and KITP Conference \citep{KITP22}.

\bibliography{bibfile}

\clearpage
\newpage

\appendix

\section{High-Energy Neutrino Fluences from Interacting SNe}\label{flux}
\begin{figure}[h!]
    \centering
    \includegraphics[width=0.5\columnwidth]{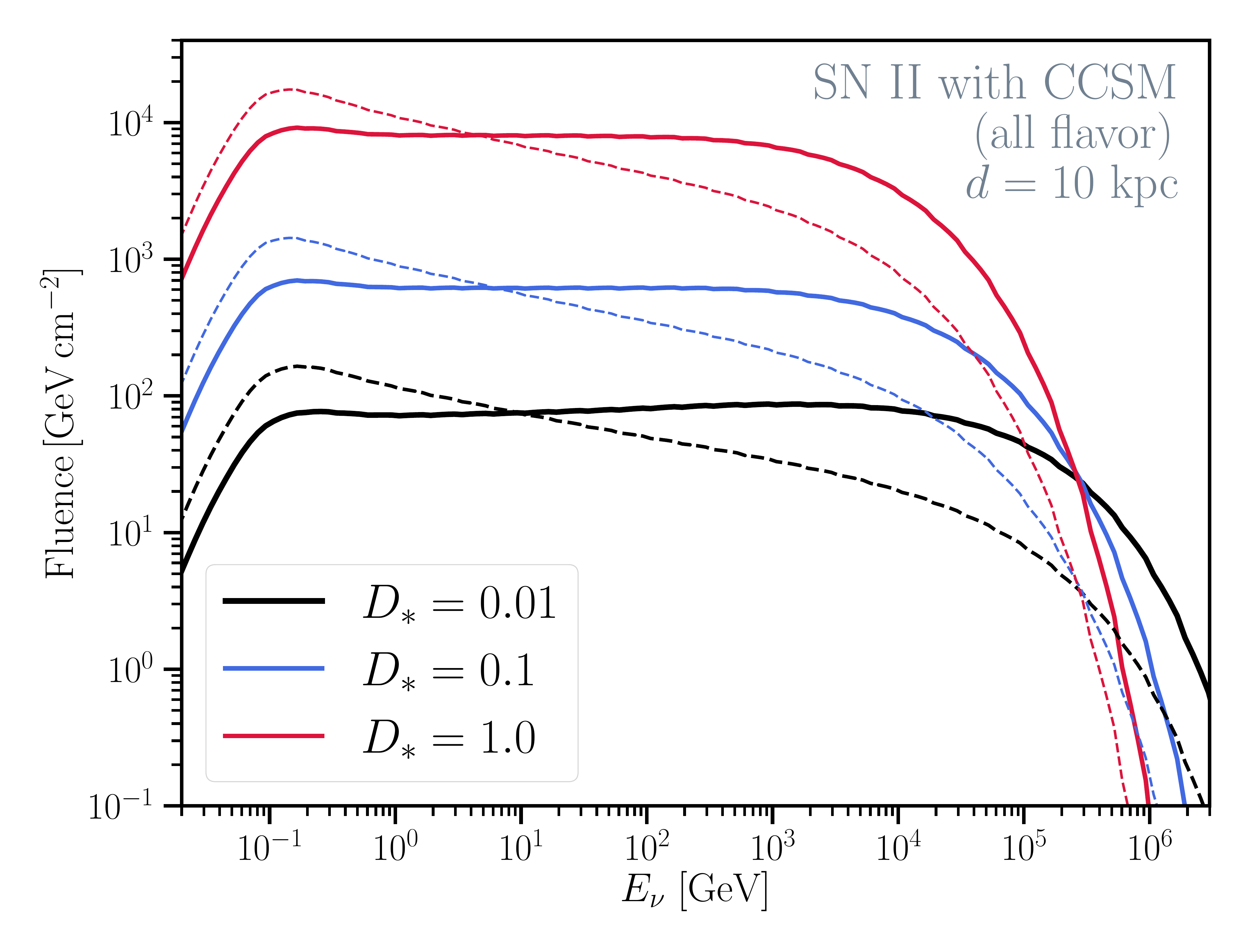}
    \caption{Time integrated high-energy neutrino fluxes for neutrino emission from SNe II with CCSM at 10 kpc, for $D_*=1$ (red), 0.1 (blue), and 0.01 (black). The latter is the $D_*$ value that was used in \citet{Murase:2017pfe}. We show the neutrino fluences for spectral indices, $s_{\rm cr}=2.0$ (solid) and $s_{\rm cr}=2.2$ (dashed).}
    \label{fig:nu_fluence_iip}
\end{figure}

\begin{figure*}[t!]
    \centering
    \includegraphics[width=0.49\columnwidth]{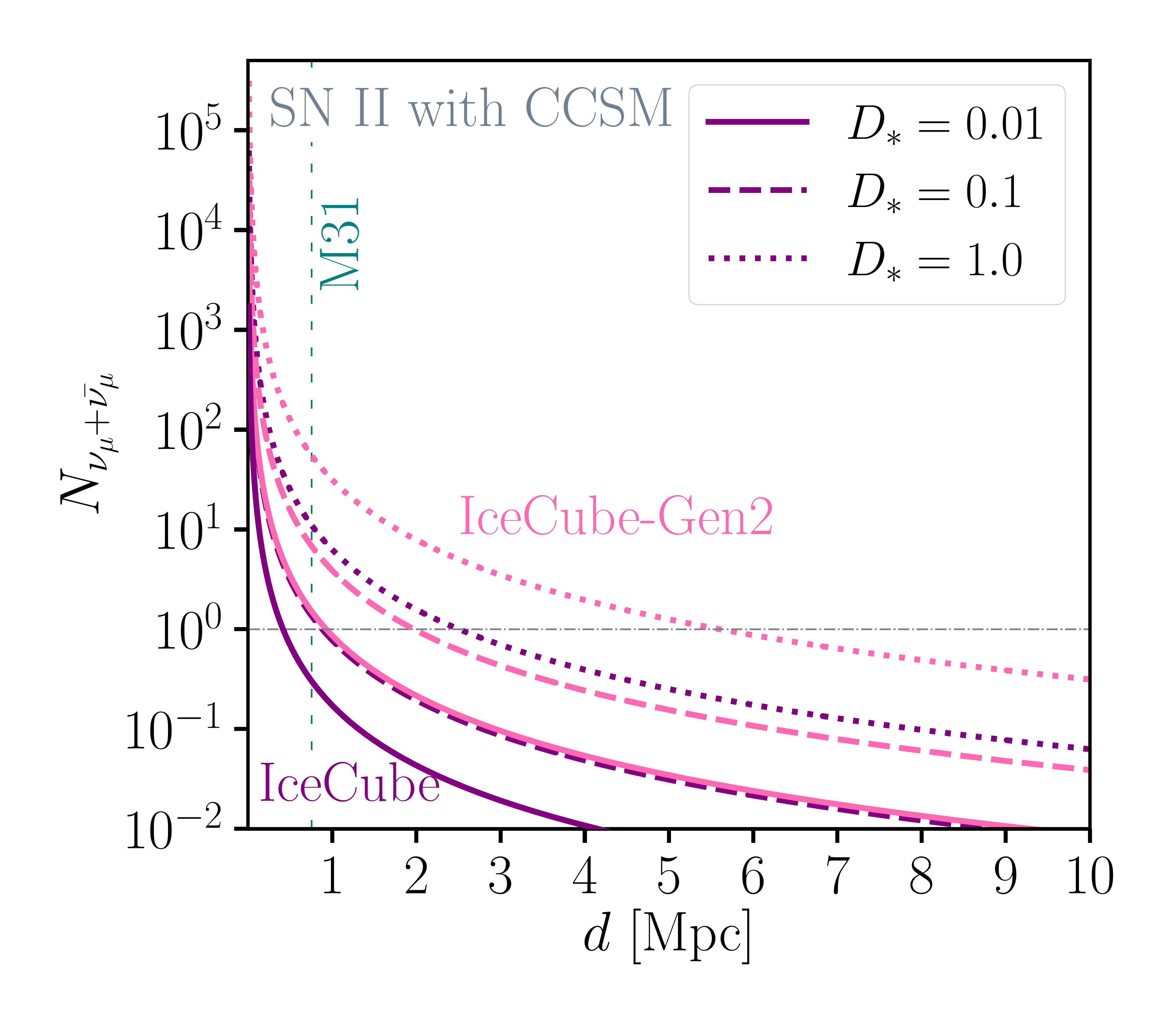}
    \includegraphics[width=0.49\columnwidth]{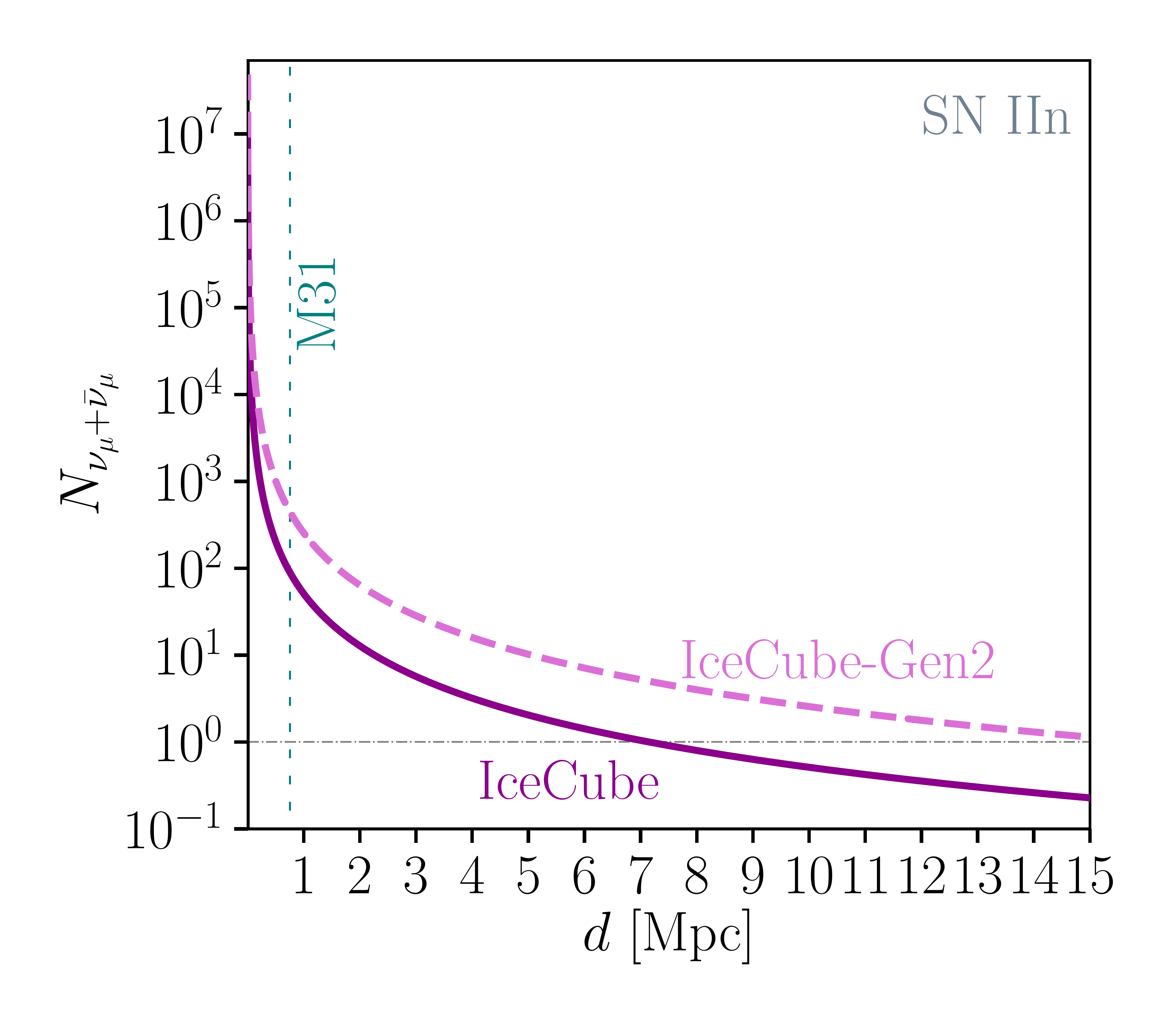}
    \caption{Left: expected neutrino signal events from SNe II with CCSM in IceCube, assuming that the source resides in the Northern sky. Here, we showcase the expected number of neutrino events between 100 GeV and 10 PeV for 3 different values of $D_*$ and $s_{\rm cr}=2.0$. Right: expected neutrino signal event from SN IIn in IceCube, assuming that the source resides in the Northern sky.}
    \label{fig:events-sn-ii}
\end{figure*}

High-energy neutrinos are produced as ions are accelerated through a strong shock formed by the interaction with CSM. Pions and kaons are produced in inelastic $pp$ interactions and promptly decay into neutrinos and gamma rays. 
Details regarding the dynamics and meson production efficiency are presented in \citet{Murase:2017pfe}, which first investigated various classes of SNe including II-P SNe with CCSM as potential high-energy neutrino sources. 
Here, we work in the same framework, assuming the wind profile for CSM, in which not only SN dynamics but also various microphysical processes are taken into account. We incorporate information from \textsc{Geant 4}~\citep{GEANT4:2002zbu}, which works better for lower energies as in \cite{Murase:2013hh}. 

For ordinary SNe II, we estimate high-energy neutrino fluences, taking into account uncertainties in the mass-loss rate based on recent optical and infrared observations. 
\citet{Murase:2017pfe} assumed $D_*=0.01$ and $R_w= 4 \times 10^{14}$ cm, motivated by early spectroscopic observations of SNe like SN 2013fs~\citep{Yaron:2017umb,Bullivant:2018tru,Chugai:2020fml}. 
On the other hand, more recent studies~\citep{2020MNRAS.496.1325B,Terreran:2021hfc,Chugai:2022fcz,WH15} allow larger values of the mass loss rate, so we consider uncertainties of $D_*$ from 0.01 to 1.0. 
For the latter, we take $R_w= 10^{15}$~cm to consider the optimistic case. Time-integrated high-energy neutrino fluxes for SNe II with CCSM are shown in Fig.~\ref{fig:nu_fluence_iip}. The assumed integrated time for each scenario is set by the optimal time~\citep[see][for details]{Murase:2017pfe}, which is $t_{\rm max} = 10^{5.8}$ s for $D_*=0.01$, $10^6$ s for $D_*=0.1$, and finally $10^{6.6}$ s for $D_*=1$.
The neutrino fluence is directly correlated to $D_*$. Given that the system is calorimetric, the estimated neutrino fluence may be increased by $\sim 2$ orders of magnitude compared to the conservative estimate based on SN 2013fs.  
For SNe IIn, we adapt the same parameter set as in \citet{Murase:2017pfe}, which corresponds to $D_*=1$ and $R_w=10^{16}$~cm~\citep[see also][for other examples including SN 2010jl and SN 2014C]{Murase:2010cu,Murase:2018okz}. 

For the calculations of neutrino fluxes employed in this study, we assume that the CR spectrum follows a power law with a spectral index of $s_{\rm cr}\sim2$ and CRs carry 10\% of the kinetic luminosity at the shock. CRs are assumed to generate via the DSA mechanism \citep{1977ICRC...11..132A,1977DoSSR.234.1306K,1978MNRAS.182..147B,1978ApJ...221L..29B}. In DSA, particles are accelerated via scatterings with plasma waves existing in both upstream and downstream of shocks. DSA is widely accepted as the principle mechanism for the acceleration of CR ions to PeV energies, especially in SN remnants, which is supported by plasma particle-in-cell simulations and observational developments~\citep[e.g.,][]{Schure:2012du,Caprioli:2013dca,Caprioli:2019xhb} for more details.. The CR spectrum and subsequently neutrino spectrum may be somewhat stepper. Both cases of $s_{\rm cr}=2.0$ and $s_{\rm cr}=2.2$ are shown in Fig.~\ref{fig:nu_fluence_iip}

\section{Expected Number of Events} \label{sec:a2}
\begin{figure*}[t]
    \includegraphics[width=0.49\columnwidth]{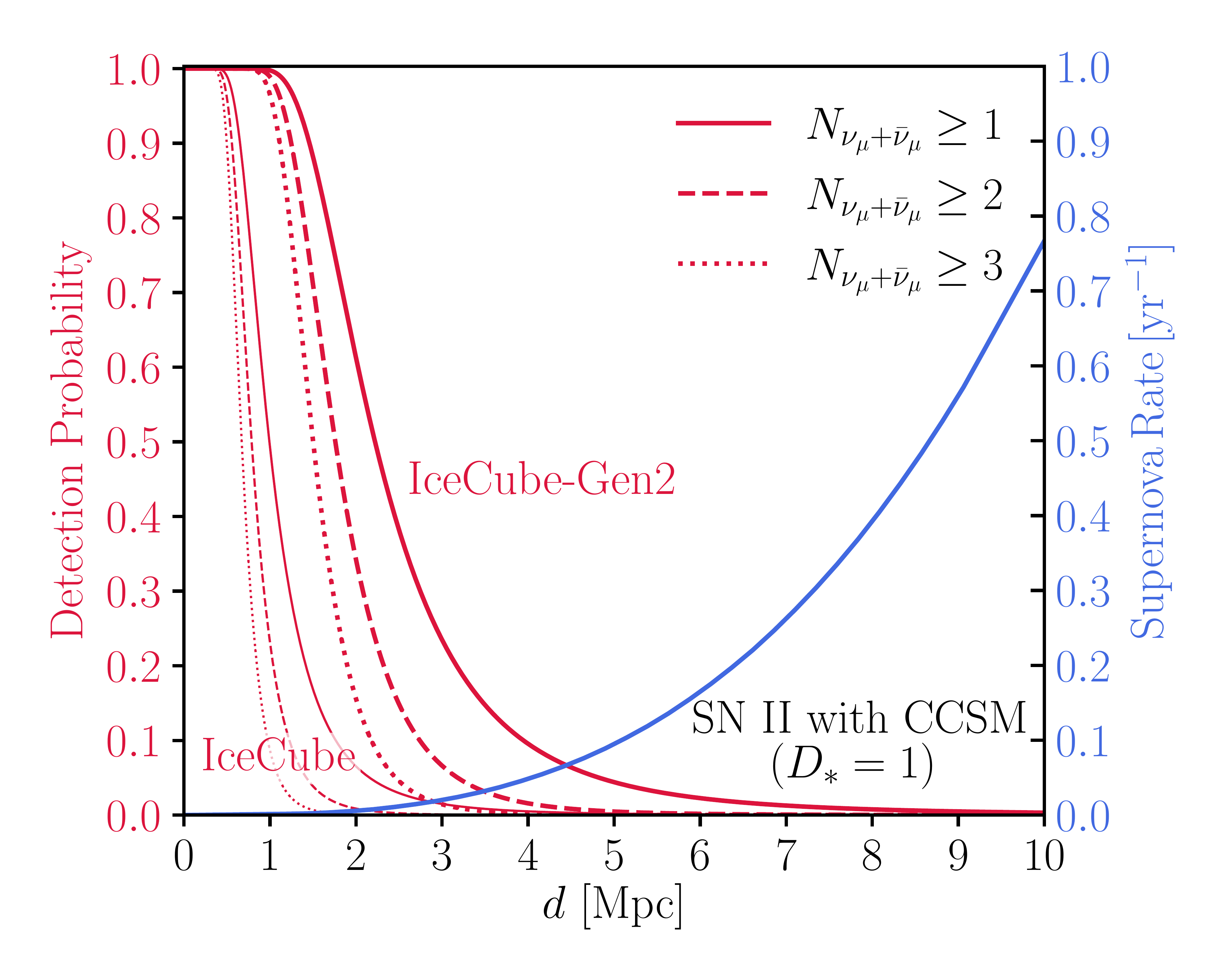}
    \includegraphics[width=0.47\columnwidth]{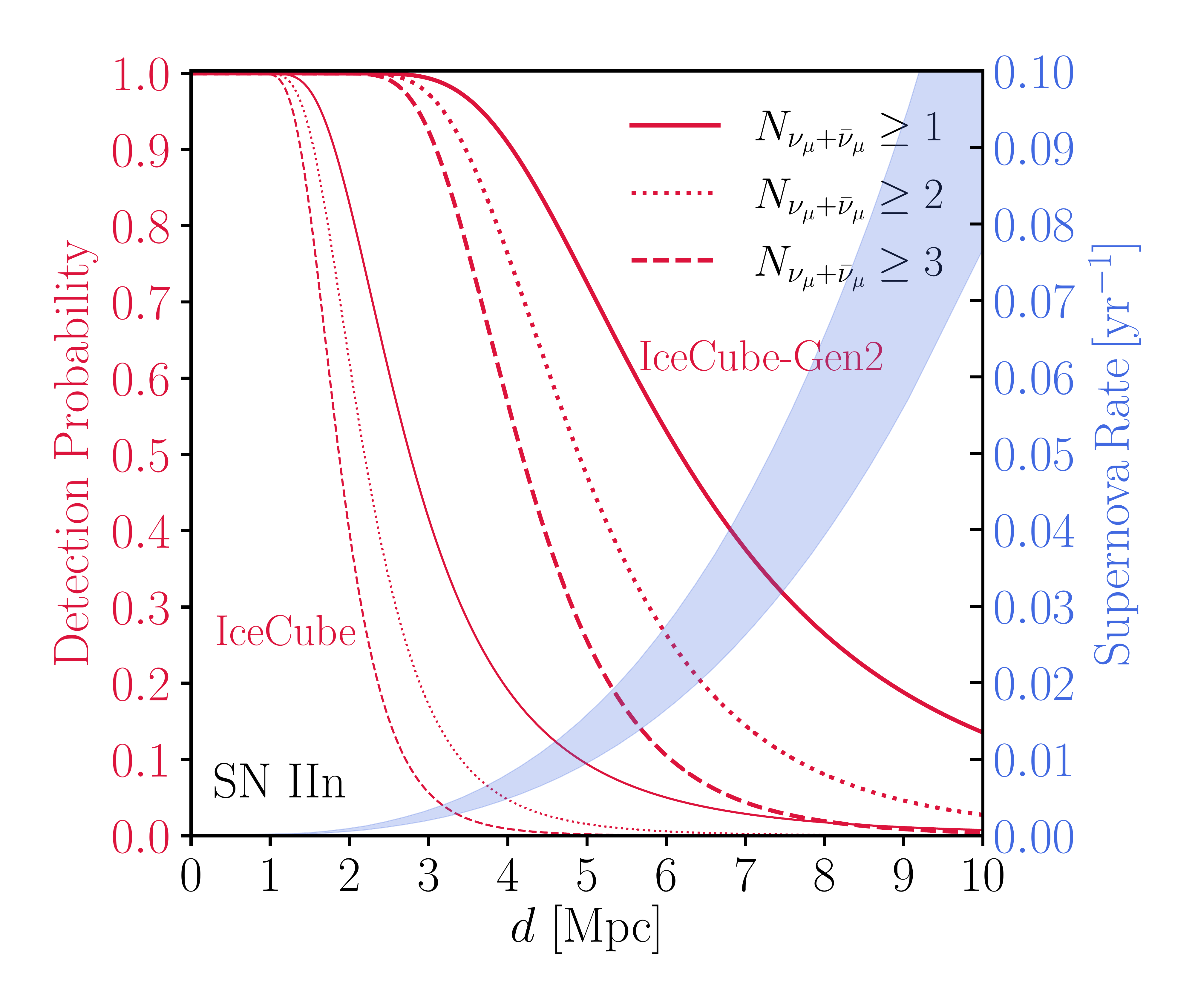}
    \caption{Detection probability for high-energy neutrinos from CCSNe in IceCube and IceCube-Gen2, assuming the CR spectral index $s_{\rm cr}=2.2$. Left: detection probability of at least one (solid), doublet (dotted), and triplet (dashed) of events from SNe II with CCSM as a function of distance for $D_*=1$. 
    Right: same as the left panel for SNe IIn. The expected rate of each SNe type is also shown in blue.}
    \label{fig:sn-iin-prospects}
\end{figure*} 

In order to evaluate the number of muon neutrino events from CCSNe in each detector, we utilize the publicly available effective areas for each detector. The expected number of muon neutrino and antineutrino events for Type II SNe with CCSM and IIn SNe are found by integrating the product of the time-integrated neutrino flux and effective area over the energy range of 100 GeV and 10 PeV. 
For IceCube, we incorporate the point source effective area~\citep{Aartsen:2016oji} for the calculation of the number of events from SNe in the local galaxies. This helps us include the Earth absorption and zenith (declination) dependence of the effective area. To calculate the generic prospects for sources in the Northern hemisphere, we use the zenith-averaged effective area as reported in \cite{Stettner:2019tok}. We should note that while this effective area is slightly larger than the point source selection, the estimated number of events is in good accordance with the previous estimate~\citep{Murase:2017pfe}. 
We scale the IceCube effective area by a factor of 5 for the expected number of events for IceCube-Gen2. We showcase the expected number of muon neutrino events in IceCube and IceCube-Gen2 for SNe II with CCSM and SNe IIn in Fig.~\ref{fig:events-sn-ii}.

For KM3NeT, we incorporate the average effective area for upgoing events according to Fig.~19 of \cite{Adrian-Martinez:2016fdl}. We do not take into account the trigger efficiency for the points-source searches. To take into account the fraction of the duration for which the source is below the horizon for KM3NeT, we rescale the number of events by the visibility of the sources, given by their declination as is defined in Fig.~37 of \cite{Adrian-Martinez:2016fdl}. In the absence of a publicly available muon effective area for GVD, we assume an IceCube-like area and take into account the visibility for particular declination of the local galaxies. 

\begin{figure*}[tb]
    \includegraphics[width=0.49\columnwidth]{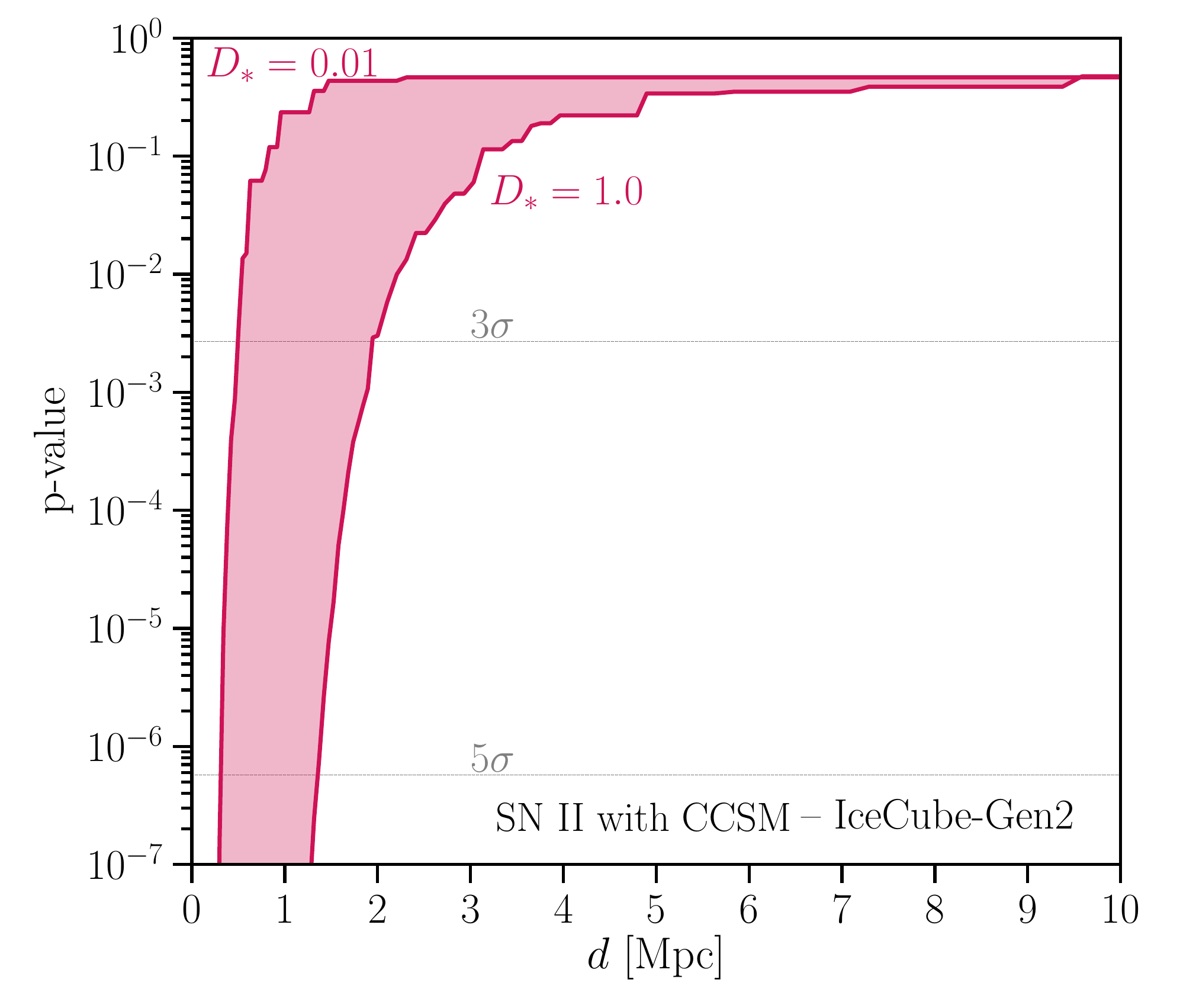}
    \includegraphics[width=0.49\columnwidth]{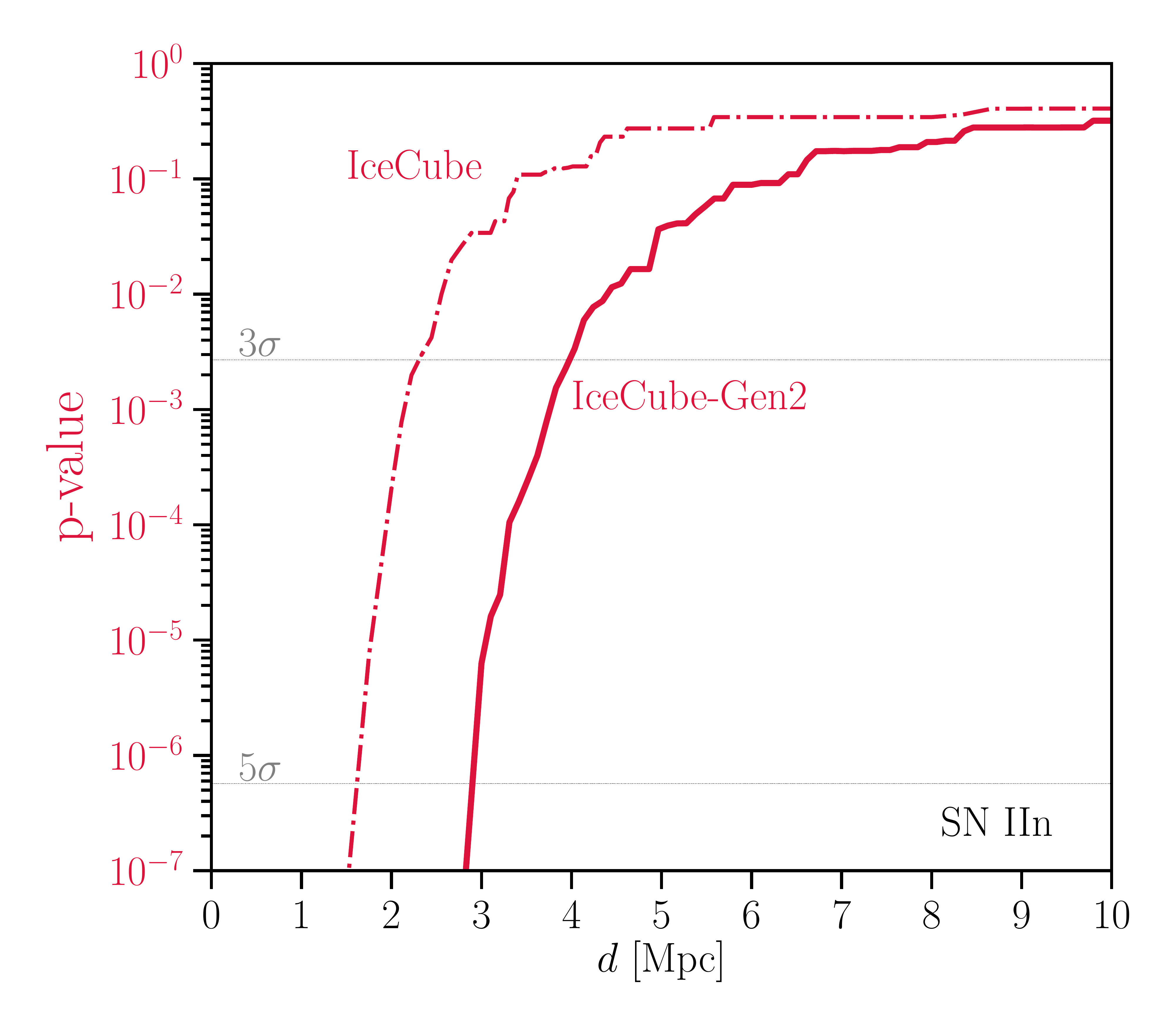}
    \caption{Left: prospects for the identification of SNe II with CCSM in IceCube-Gen2 at different distances. The shaded band shows the p-values for $D_*$ ranging from 0.01 to 1. 
    Right: same as left but for SN IIn in IceCube (dashed) and IceCube-Gen2 (solid).}
    \label{fig:gen2-pval}
\end{figure*}

We calculate the expected background of atmospheric neutrinos is estimated from the using the zenith dependent atmospheric neutrino flux reported by \cite{Honda:2006qj}, assuming an angular resolution of 0.4$^\circ$ uniformly for all experiments considered in this study.

We estimate the statistical significance for the identification of a SN by using the analytic expression introduced by \cite{CMS-NOTE-2011-005}, which has been previously employed to evaluate the prospects for observation of high-energy neutrino sources~\citep[see, e.g.,][]{Gonzalez-Garcia:2013iha,Halzen:2016seh}. In this method, the p-value of identifying signal events from a source over the background is given by
\begin{equation}
p_{\rm value}=\frac{1}{2}\left[1-{\rm{erf}} 
\left( \sqrt{q_0^{\rm obs}/2} \right) \right]\,,
\end{equation}
where $q_0^{\rm obs}$ is defined as 
\begin{equation}
q_0^{\rm obs} \equiv -2 \ln \mathcal{L}_{b,D} = 2 \sum_i \left( Y_{b,i} - N_{D,i} + N_{D,i} \ln \left( \frac{N_{D,i}}{Y_{b,i}}\right)\right)\, ,
\end{equation}
with $i$ runs over the different energy bins. Here, $Y_{b,i}$ is the theoretical expectation for the background hypothesis, while $N_{D,i}$ is the estimated signal generated as the median of events Poisson-distributed around the signal plus background for given time windows. The background expectation here is the number of atmospheric neutrinos in the solid angle set by the angular resolution. This solid angle corresponds to roughly 72\% of the signal events from the source~\citep[see][for a discussion]{Alexandreas:1992ek}. 

\begin{figure}[th]
    \centering
    \includegraphics[width=0.5\columnwidth]{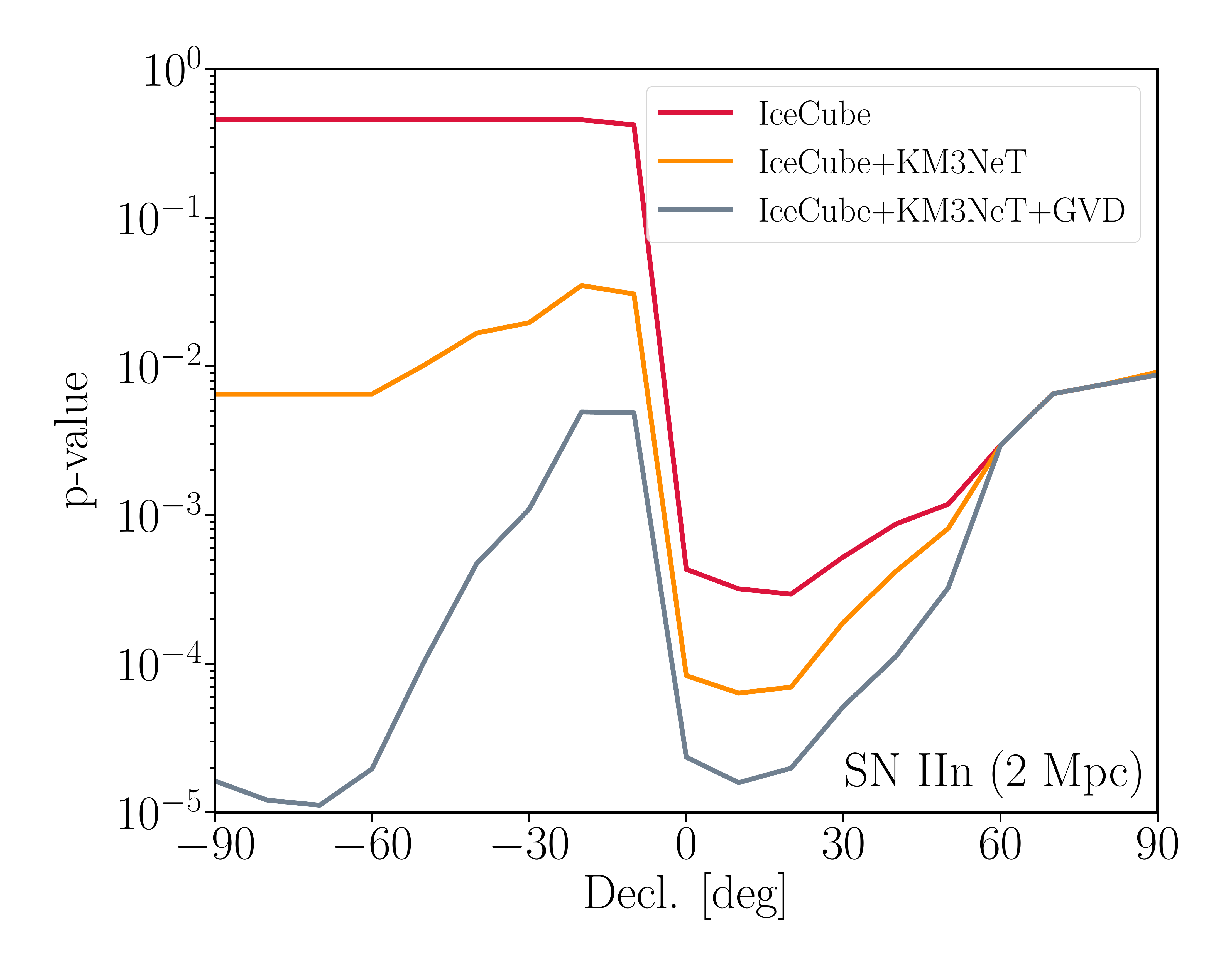}
    \caption{Projected p-values for the detection of neutrinos from SN IIn at a distance of 2 Mpc in IceCube (red), IceCube and KM3NeT (orange), and a combination of IceCube, KM3NeT, and GVD.}
    \label{fig:pvalzenith}
\end{figure}

Fig.~\ref{fig:gen2-pval} (right panel) shows the projected p-values as a function of distance for SNe IIn for IceCube and IceCube-Gen2. Thanks to the high level of fluences of SNe IIn, the 3$\sigma$ horizon is extended to near 4 Mpc. In Fig.~\ref{fig:gen2-pval} (left panel), we also show the projected p-values for SNe II with CCSM as a function of distance. The shaded region shows the expectations for conservative ($D_*=0.01)$ and optimistic ($D_*=1$). In addition, to further demonstrate the power of joint searches with the network of neutrino telescopes, we showcase the expected p-values of detecting neutrinos from a SN IIn at 2 Mpc for the combination of IceCube and soon-to-be operational telescopes, KM3NeT and GVD, in Fig.~\ref{fig:pvalzenith}.

\section{Application to SN 2023ixf}\label{ixf}

\begin{figure}[th]
    \centering
    \includegraphics[width=0.5\columnwidth]{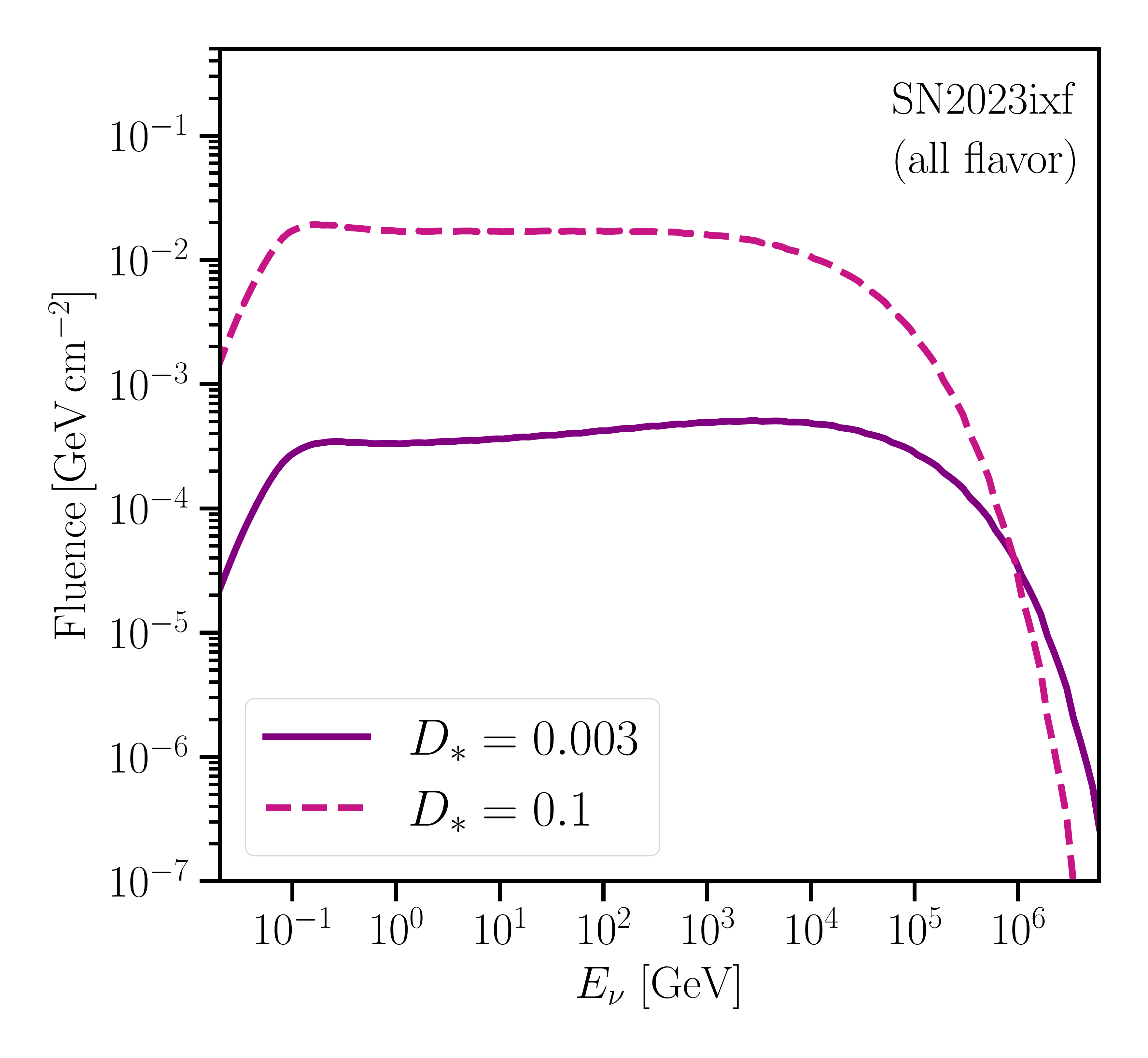}
    \caption{Neutrino fluences for SN 2023ixf, identified in a relatively nearby galaxy M101. The solid (dashed) line shows the fluence for $D_* = 0.003\, (0.1)$, where $s_{\rm cr}=2.0$ is used.}
    \label{fig:sn2023ixf}
\end{figure}

A recent, relatively nearby, core collapse SN observed in M101 \citep{Perley2023} brought additional interest and attention to the potential neutrino signals and motivated multi-messenger searches, especially in light of indications of dense circumstellar medium \citep{Yamanaka2023}. The interacting nature of the transient event was established based on multi-wavelength observations. Here, we present the expected neutrino fluence from this source. We estimate the neutrino spectrum for two values of $D_*$, 0.003 and 0.1. The former is deduced from X-ray observations \citep{Grefenstette:2023dka} and the latter from optical measurements \citep{Jacobson-Galan:2023ohh}. We show the expected neutrino fluence for both cases in Fig.~\ref{fig:sn2023ixf}. These predicted fluences yield approximately 0.0007 and 0.02 events for $D_* = 0.003$ and 0.1, respectively. This is compatible with the results reported by the IceCube fast response analysis~\citep{IceCubeSN2023}.

\end{document}